\input epsf
\documentstyle[prd,floats,aps]{revtex}
\begin{document}
\title{Variational derivation of exact skein relations from 
Chern--Simons theories}

\author{Rodolfo Gambini\\{\em Instituto de F\'{\i}sica, Facultad de
Ciencias,\\Tristan Narvaja 1674, Montevideo, Uruguay}}

\author{Jorge Pullin\\  {\em Center for Gravitational Physics and Geometry}\\
{\em Department of Physics, 104 Davey Lab,}\\ {\em The Pennsylvania State 
University,}\\
{\em University Park, PA 16802}}

\maketitle
\begin{abstract}
The expectation value of a Wilson loop in a Chern--Simons theory is a
knot invariant. Its skein relations have been derived in a variety of
ways, including variational methods in which small deformations of the
loop are made and the changes evaluated.  The latter method only
allowed to obtain approximate expressions for the skein relations.  We
present a generalization of this idea that allows to compute the exact
form of the skein relations. Moreover, it requires to generalize the
resulting knot invariants to intersecting knots and links in a manner
consistent with the Mandelstam identities satisfied by the Wilson
loops. This allows for the first time to derive the full expression
for knot invariants that are suitable candidates for quantum states of
gravity (and supergravity) in the loop representation. The new
approach leads to several new insights in intersecting knot theory, in
particular the role of non-planar intersections and intersections with
kinks.
\end{abstract}

\vspace{-10.5cm} 
\begin{flushright}
\baselineskip=15pt
CGPG-96/2-3  \\
gr-qc/9602165\\
\end{flushright}
\vspace{9.5cm}

\section{Introduction}

Witten \cite{Wi89} realized some years ago that the expectation value of
a Wilson loop $W(\gamma)$ in a Chern--Simons theory was a knot
invariant. This follows from the fact that Chern--Simons theories are
diffeomorphism invariant and that the Wilson loops are observables for
such theories, having therefore diffeomorphism invariant expectation
values. The resulting knot invariant is the Kauffman bracket
\cite{KaBr} for the case of an $SU(2)$ Chern--Simons theory and for
the case of $SU(N)$ is a regular isotopic polynomial associated with
the HOMFLY \cite{HOMFLY} polynomial.  These results were derived based
on the calculations of Moore and Seiberg \cite{MoSe} for the
monodromies of rational conformal field theories. The knowledge of the
Yang-Baxter relations satisfied by the monodromies translates immediately
into skein relations for the polynomial in question.

Independently, Smolin \cite{Sm} and later Cotta-Ramusino, Guadagnini,
Martellini and Mintchev \cite{CoGuMaMi} noted that a simpler heuristic
derivation of the skein relations was possible. The idea is similar to
the Makeenko-Migdal \cite{MaMi} approach to Yang--Mills theories. It
is based on studying the changes in the expectation value of the
Wilson loop when one performs small deformations. This calculation can
be done explicitly to first order in the deformation. The results can
be interpreted as skein relations to first order in the inverse
coupling constant of the theory, which is tantamount to determining
the knot polynomial to first order.

The latter method is quick and computationally efficient, and has a
simple generalization to the case of intersecting loops
\cite{BrGaPunpb,KaBa}. The main drawback, especially in the case where
the result is not known by other methods, is that one only gets the
skein relations to first order in the inverse of the coupling constant of
the theory. It is therefore of interest to find a suitable
generalization that would yield the exact skein relations to all
orders. This is the main purpose of this paper.

On the other hand, the subject of intersecting knot invariants has
received little attention and is of paramount importance for the
construction of quantum states of gravity in the loop representation
\cite{GaTr86,RoSm90}.  In this approach, based on the canonical
quantization of general relativity in terms of Ashtekar variables
\cite{As86}, wavefunctions are knot invariants due to the diffeomorphism
symmetry of general relativity \cite{RoSm88}. The Hamiltonian
constraint has only a non-trivial action at intersections
\cite{JaSm}. Only intersecting knots are associated with non-degenerate
spacetimes \cite{BrPu}. Whenever one generalizes an invariant of smooth
loops to take values on intersecting loops there is generically
freedom in how the invariant is defined, as long as it is compatible
with the Reidemeister moves. However, in the case of quantum gravity,
wavefunctions have to be compatible with a set of constraints among
functions of loops known as the Mandelstam identities. These
identities naturally involve intersecting loops and severely limit the
possible generalizations of invariants to intersections. We will show
in this paper how to generalize the invariants stemming from
Chern--Simons theory to be compatible in an exact way with the
Mandelstam identities. This in particular also defines the values of
the invariants for multicomponent links. This is of particular
relevance for quantum gravity since it is known that the exponential
of the Chern--Simons form built from the Ashtekar connection is an
exact solution to all the constraints of quantum gravity
\cite{Ko,BrGaPunpb}. If one wishes to find the counterpart of this state
in the loop representation one ends up computing exactly the same
integral as the expectation value of a Wilson loop in a Chern--Simons
theory. 

One additional motivation for the construction we present is that the
Chern--Simons state not only arises in canonically quantized vacuum
general relativity but also in other contexts, like
Einstein--Yang--Mills theories \cite{GaPu} and supergravity
\cite{ArUgGaObPu}. In these cases the (super)gauge group of the
associated Chern--Simons theory differs from that of gravity and
therefore so do the resulting invariants. In the particular case of
supergravity the resulting invariant had not been computed by other
means, and turns out to be associated with the Dubrovnik--Kauffman
polynomial \cite{DuKa}.

A first attempt to obtaining a finite prescription from variational
calculations was made by Br\"ugmann \cite{Br}. In particular, the idea of
exponentiating the infinitesimal transformation was presented
there. Because of ambiguities in the formulation presented in that
paper, one could only check, given the exact skein relations, that
they were compatible with the formulation.  Here we add two key
elements that make the construction a well defined prescription:
on the one hand we offer a justification of why can the infinitesimal
results be exponentiated; on the other hand we make crucial use of the
Mandelstam identities to uniquely fix a prescription for the
exponentiation. The prescription now  allows,
in a case where one does not know the result beforehand, to compute
the resulting polynomial.

The organization of this article is as follows. In the next section we
will discuss, using the non-Abelian Stokes theorem, how to perform a
finite deformation of the expectation value of the Wilson loop.In
section III we discuss the deformation of twists and kinks and in
section IV of planar and non-planar intersections. We discuss the
implications of the results in section V.

\section{The non-Abelian Stokes theorem and finite deformations of loops}

One is interested in establishing skein relations for the following 
function of a loop,

\begin{equation}
<W(\gamma)> = \int dA \exp\left({i k\over 4 \pi} S_{CS}\right) W(\gamma)
\end{equation}
where
\begin{equation}
W(\gamma) = {\rm Tr}\left[ {\rm P}\exp\left( i\oint dy^a A_{a}(y)\right)\right]
\end{equation}
where $\gamma$ is a closed curve in a three manifold, $A_a$ is a connection
in a semi-simple Lie algebra and $S_{CS}$ is the action of a 
Chern--Simons theory,

\begin{equation}
S_{CS} = \int d^3x\,
 \epsilon^{abc} {\rm Tr}[A_a \partial_b A_c -{2 \over 3} i 
A_a A_b A_c]. 
\end{equation}

\begin{figure}[t]
\hspace{3cm}\epsfxsize=300pt \epsfbox{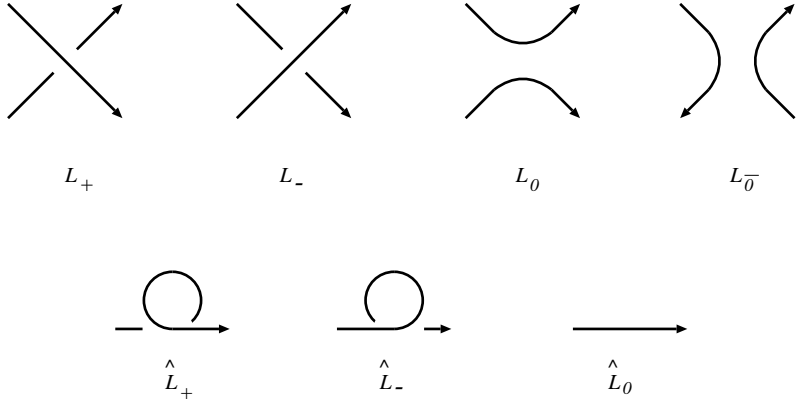}
\caption{The different crossings involved in the skein relations for 
invariants of non-intersecting loops}
\label{skeinrelfig}
\end{figure}

Finding skein relations involves relating the value of the
$<W(\gamma)>$ for different loops. These loops are determined by the
replacement of over-crossings by under-crossings in a planar
projection. Skein relations also arise between a loop with and without
a twist,  over-crossings and intersections.  The usual
notation for the several types of crossings is shown in figure
\ref{skeinrelfig}. We will see later on that the consideration of
invariants with intersections requires several other crossings. For
instance, the skein relations that define the Kauffman bracket knot
polynomial on loops without intersections are,

\begin{eqnarray}
K(\hat{L}_+) &=& q^{3/4} K(\hat{L}_0)\label{elemas}\\
K(\hat{L}_-) &=& q^{-3/4} K(\hat{L}_0)\label{elemenos}\\
q^{1/4} K(L_+) - q^{-1/4} K(L_-) &=& (q^{1/2}-q^{-1/2}) K(L_0)
\label{elemasmenos}\\
K({\rm unknot}) = 1.
\end{eqnarray}

With these relations the polynomial is completely characterized for
any link. What Witten \cite{Wi89} showed using conformal field theory 
techniques is that ${1\over 2}
<W(\gamma)>$ satisfies the above relations. Here we will show it by
performing directly deformations of the loops on the expression
of the expectation value.

In order to do this, we need to study deformations of Wilson loops. 
Wilson loops are traces of holonomies. It turns out that the information
needed to compute a holonomy is less than that present in
a closed curve. Several closed curves yield the same holonomy. 
In this paper we will use the word ``loop'' to denote the equivalence
class of curves that yield the same holonomy for all connections
\footnote{Other authors call these objects ``hoops'' to denote 
``holonomic loops'' \cite{AsLe}.}. Loops form a group structure
called the group of loops \cite{GaPu}. 

\begin{figure}[h]
\hspace{3cm}\epsfxsize=300pt \epsfbox{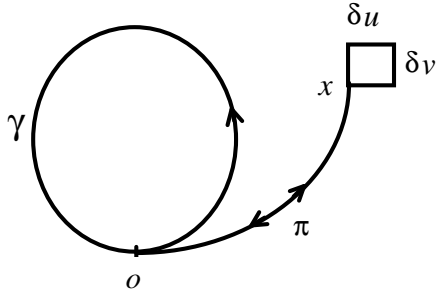}
\vspace{-9cm}
\caption{The infinitesimal loop that defines the loop derivative.}
\label{loopder}
\end{figure}

It is well known that if one adds to a loop another loop of 
infinitesimal area, the change in the Wilson loop can be coded in 
terms of an infinitesimal operator in loop space called the loop
derivative \cite{GaPu}. This is shown in figure \ref{loopder}. The
concrete expression is,

\begin{equation}
W(\gamma\circ \delta \gamma) = (1+{1\over 2} \sigma^{ab} 
\Delta_{ab}(\pi_o^x)) W(\gamma)
\end{equation}
where $\circ$ denotes composition of loops, $\sigma^{ab}$ is the
infinitesimal element of area of the the loop $\delta \gamma$ and
$\pi_o^x$ is an open path connecting the basepoint of the loop
$\gamma$ to the point $x$ at which one adds the infinitesimal loop
with infinitesimal element of area $\sigma^{ab}$. Notice that the loop
derivative $\Delta_{ab}$ depends on the path used to compute it and 
we denote so in its expression. The loop derivative is related to the
infinitesimal generators of the group of loops \cite{GaPu}.

\begin{figure}[t]
\vspace{-0.9 cm}
\hspace{3 cm}\epsfxsize=300pt \epsfbox{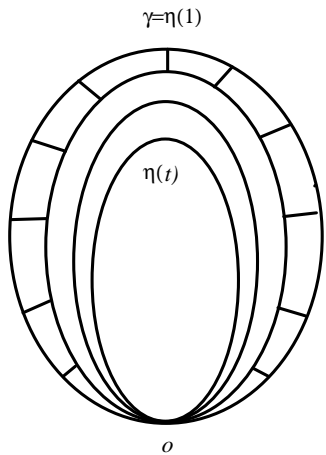}
\vspace{-8cm}
\caption{The construction of a finite loop from the loop derivative. The
curves $\delta \eta_i$ are determined by two elements in the family $\eta(t)$.}
\label{stokes}
\end{figure}

One can write \cite{GaPu} an expression for an operator $U(\gamma)$ that
adds a {\em finite} loop $\gamma$ to a Wilson loop in terms of the loop
derivative.  Let $\gamma(s)$ be a parameterized curve belonging to the
equivalence class defining the finite loop $\gamma$ with $s\in [0,1]$.
Consider a one-parameter family of parameterized loops $\eta(s,t)$
interpolating smoothly between $\gamma(s)$ and the identity loop, such
that $\eta(s,0)$ is in the equivalence class of the identity loop and
$\eta(s,1)=\gamma(s)$. Consider the curves $\eta(s,1)$ ($=\gamma(s)$)
and $\eta(s,1-\epsilon)$. The two curves are drawn in figure
\ref{stokes} and differ by an infinitesimal element of area.
The whole purpose of our construction will be to cover the infinitesimal area
separating the two mentioned curves with a ``checkerboard'' of
infinitesimal closed curves such that along each of them one can
define a loop derivative. One can therefore express the curve
$\gamma(s)$ as
\begin{equation}
\gamma(s) = \lim_{n\rightarrow\infty} \eta(s,1-\epsilon)\circ \delta
\eta_1 \circ \cdots \circ \delta \eta_n,
\end{equation}
where the $\delta \eta_i$ are shown in figure \ref{stokes}.
Analytically, in terms of differential operators on functions of
loops we can write\footnote{We drop the $s$ dependence of $\eta$ where
it is not relevant.}
\begin{eqnarray}
\Psi(\eta(1)) =&& \Psi(\eta(1-\epsilon))  \nonumber\\&&
+\epsilon \oint_0^1 ds
\dot{\eta}^a(1-\epsilon,s) {\eta'}^b(1-\epsilon,s)
\Delta_{ab}(\eta(1-\epsilon)_o^s) \Psi(\eta(1-\epsilon)),\nonumber\\
\end{eqnarray}
where $\dot{\eta}(t,s)\equiv d\, \eta(t,s)/dt$ and $\eta'(t,s)
\equiv d\, \eta(t,s)/ds$.
It is
immediate to proceed from $\eta(1-\epsilon,s)$ inwards just by
repeating the same construction, and so continuing until the final
curve is the identity. The end result is
\begin{equation}
\Psi(\gamma) = {\rm T}\exp\left(\int_0^1 dt \oint_0^1 ds
\dot{\eta}^a(t,s) {\eta'}^b(t,s)
\Delta_{ab}(\eta(t)_o^s) \right)\Psi(\eta(0))\equiv U(\gamma) \Psi(\eta(0))
\end{equation}
where the outer integral is ordered in $t$ (T-ordered). This result is
the loop version of the non-Abelian Stokes theorem of gauge theories
\cite{Ar80} and it shows that the loop derivative is a generator of
loop space, i.e., it allows us to generate any finite loop homotopic to
the identity. Due to the properties of the group of loops \cite{GaPu}, 
the construction is {\em independent} of the
particular family of loops used to go from the identity element to the 
final loop $\gamma$. 

It is useful to rewrite the expression of the operator $U$ as,
\begin{equation}
U(\gamma) = {\rm T} \exp \int_0^1 dt \delta(t)
\end{equation}
where
\begin{equation}
\delta(t) = \oint_0^1 ds
\dot{\eta}^a(t,s) {\eta'}^b(t,s)
\Delta_{ab}(\eta(t)_o^s). 
\end{equation}

The operator $\delta$ is closely related to the unparameterized
``connection derivative'' \cite{GaPu}.

We would now like to apply the above deformation to a portion of a loop
in order to construct the elements that appear in the skein relations. 

\section{Skein relations associated to twists and kinks}
\subsection{Twists}
\begin{figure}
\hspace{7 cm}\epsfxsize=130pt \epsfbox{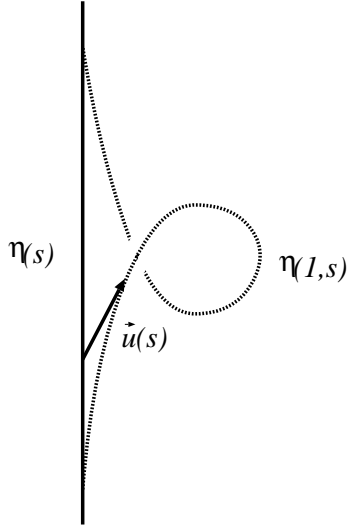}
\caption{Adding a twist to a loop using the finite deformation operator}
\label{deforfig}
\end{figure}

We start by the
simplest skein relations, those that relate the value of the invariant
with and without a twist. Starting from a regular portion of a loop
$\eta_o^x$ going from the origin to $x$, one can add a twist by 
considering a family of loops of the form,

\begin{equation}
\eta^a(s,t) = \eta^a(s)+ t u^a(s)
\end{equation}
where $u^a(s)$ is a vector along the loop that materializes the deformation
shown in figure \ref{deforfig}.

To compute the deformation, we first evaluate the action of the
loop derivative on the expectation value of the Wilson loop. 

\begin{equation}
\sigma^{ab} \Delta_{ab}(\eta(t)_o^x) <W(\gamma)> = 
i \int dA \sigma^{ab} F_{ab}^k(x) {\rm Tr}(\tau^k H(\eta(t)_x^o\circ
\gamma\circ \eta(t)_o^x)) e^{{i k\over 4 \pi} S_{CS}}
\label{actionofloopder}
\end{equation}
where we have taken into account that the deformation is along one
element of the the family of loops $\eta(t)$. $\tau^k$ are the generators
of the algebra $su(2)$. In this calculation we have assumed that the loop
derivative acts on $<W(\gamma)>$ by simply acting inside the functional
integral $W(\gamma)$. This is a strong hypothesis, notice that the integral
$<W(\gamma)>$ is diffeomorphism invariant and therefore the limit involved
in the definition of the loop derivative is singular. We are assigning
a value to that limit by permuting the functional integral and the 
limit involved in the loop derivative.

We now use the fact that the field tensor is the dual of the magnetic
field, which can in turn be obtained through a functional derivative of
the exponential of the Chern--Simons state,

\begin{equation}
i F_{ab}^l(x) e^{{i k\over 4 \pi} S_{CS}} = 
i \epsilon_{abc} B^{c\,l} e^{{i k\over 4 \pi} S_{CS}} = 
{4 \pi \over k} 
\epsilon_{abc}{\delta \over \delta A_c^l} e^{{i k\over 4 \pi} S_{CS}}.
\end{equation}

Substituting this expression in (\ref{actionofloopder}) and integrating the
functional derivative by parts we get,
\begin{equation}
\sigma^{ab} \Delta_{ab}(\eta(t)_o^x) <W(\gamma)> = 
-{4 \pi \over k}  \int dA \sigma^{ab} \
e^{{i k\over 4 \pi} S_{CS}}
\epsilon_{abc}{\delta \over \delta A_c^l} 
{\rm Tr}(\tau^l H(\eta(t)_x^o\circ
\gamma\circ \eta(t)_o^x)) 
\end{equation}
and acting with the functional derivative on the holonomy we get,
\begin{equation}
\sigma^{ab} \Delta_{ab}(\eta(t)_o^x) <W(\gamma)> = 
-{4 \pi i\over k}  \int dA \sigma^{ab} 
e^{{i k\over 4 \pi} S_{CS}}
\epsilon_{abc} \oint_\gamma dy^c \delta^3(x-y)
{\rm Tr}\left(\tau^lH(\eta(t)_x^o)
H(\gamma_o^y)\tau^lH(\gamma_y^o)H(\eta(t)_o^x)\right).
\end{equation}

Notice that the previous expression is distributional, involving a
one-dimensional integral of a three-dimensional Dirac delta. It is 
remarkable that one can use it to obtain an expression for the 
operator $U$ that generates finite deformations. In order to see 
this, we first write an expression for the operator $\delta$ 
associated with the deformation induced by the vector field $\vec{u}(s)$,
\begin{equation}
\delta_{\vec{u}}(t) <W(\gamma)> = -{4 \pi i\over k} \int_0^1 ds u^a(s) 
({\eta'}^b(s)+t {u'}^b(s)) \epsilon_{abc} 
\oint_\gamma dy^c \delta^3(\eta(s,t)-
y) \left<{\rm Tr}\left(\tau^lH(\eta(t)_x^o)
H(\gamma_o^y)\tau^lH(\gamma_y^o)H(\eta(t)_o^x)\right)\right>.
\label{actionofdeltat}
\end{equation}

In order to construct the finite deformation operator we need to
exponentiate $\delta$. As is shown in appendix A, one can find a
regularization such that $\delta$'s at different points commute when
acting on the expectation value of the Wilson loop in Chern-Simons
theory (in general they do not commute, see \cite{GaPu}). Therefore
the $T$-ordered exponential reduces to an ordinary exponential and 
we get,
\begin{equation}
U(\gamma) =\exp\left(\int_0^1 \delta_{\vec{u}}(t)dt \right)
\end{equation}
when acting on $<W(\gamma)>$ and the integral in the exponent can be computed
explicitly,
\begin{equation}
\int_0^1 dt \delta_{\vec{u}}(t) <W(\gamma)> = 
\mp {3 i \pi\over k} <W(\gamma)> 
\end{equation}
and the sign $\mp$ corresponds to the kind of crossing generated in 
the loop by the vector field $\vec{u}$. Positive sign corresponds to 
the right hand rule. To go from \ref{actionofdeltat} to the last 
expression we have made use of the Fierz identity for $SU(2)$,
\begin{equation}
\tau^k{}^A_B \tau^k{}^C_D = {1 \over 2} \delta^A_D \delta^C_B -{1\over 4}
\delta_B^A \delta_D^C,
\label{Fierz}
\end{equation}
and we performed explicitly the three one-dimensional integrals. The
result is the normalized oriented volume subtended by the deformation.
Strictly speaking, this result implies a choice of regularization,
since the type of integral that one is left with is of the form
$\int_0^1 dt \delta(t)$ (see \cite{Br} for details). This choice in
the regularization is tantamount to introducing a new parameter in the
derivation, corresponding to the value of the above
integral. Throughout this paper we will take it to be
unity. Otherwise, it would imply a multiplicative shift in the value
of $k$.

The above expression can be summarized in terms of the notation for
skein relations we introduced before as,
\begin{eqnarray}
<W(\hat{L}_+)> &=&\exp\left(-{3 i \pi\over k} \right)<W(\hat{L}_o)> 
\label{skeintwist+}\\
<W(\hat{L}_-)> &=&\exp\left({3 i \pi\over k} \right)<W(\hat{L}_o)> 
\label{skeintwist-}
\end{eqnarray}
which coincide with the skein relations for the Kauffman bracket
(\ref{elemas},\ref{elemenos}) if one makes the identification
\begin{equation}
q=\exp(-{4 i \pi\over k}).
\end{equation}
It is well known that perturbative techniques like the ones we are
using here fail to capture the additive shift in the coupling constant 
first observed by Witten \cite{Wi89}. A discussion, and a proposal
to amend perturbative techniques to capture this effect through
a recourse to the semi-classical approximation can be found in 
Awada \cite{Aw}. We could proceed in the same fashion here and
modify the value of the coupling constant, but we will leave
it as it is for simplicity.

\subsection{Kinks}

A kink (discontinuity of the tangent) is a diffeomorphism invariant
feature of a loop. We will now show that the expectation value of
a Wilson loop in a Chern--Simons theory is not sensitive to the 
presence of kinks in the loop under the kind of regularization
we are using in our calculations. This is an important result
because it directly relates to the value of the invariant when
one has intersections with kinks, which are crucial to implement
the Mandelstam identities. We will discuss this at the end of 
this section and will see later that the choice we make for the
treatment of the kinks is central to establishing consistency.

\begin{figure}
\hspace{8 cm}\epsfxsize=80pt \epsfbox{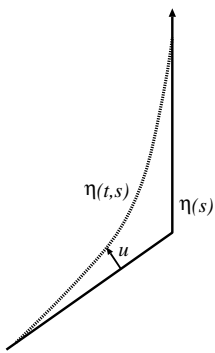}
\caption{Deforming a kink into a smooth section}
\label{kinkfig}
\end{figure}

Let us consider a loop with a kink like that shown in figure
\ref{kinkfig}. The kink defines a plane. We can deform the kink 
into a smooth section of the loop through a deformation in the plane.
The calculation is exactly the same as that of a smooth section, so we
will not repeat it here. The only precaution is to consider the
appropriate tangent vector at both sides of the kink. The reason why
we do not need more details is that since the deformation is planar,
the contribution vanishes. Basically one gets a contribution similar
to (\ref{actionofdeltat}) that has three coplanar vectors contracted
with the Levi--Civita symbol.

\begin{figure}
\hspace{5 cm}\epsfxsize=200pt \epsfbox{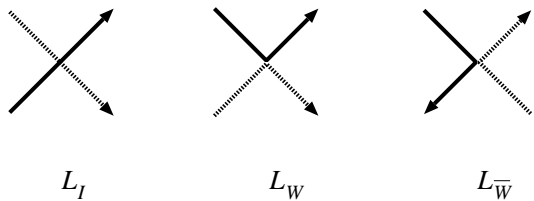}
\caption{Double intersections without and with kinks}
\label{intersecfig}
\end{figure}

The case of intersections with kinks is essentially similar. There
are two different types of (double) planar intersections with kinks,
as shown in figure \ref{intersecfig}.

\begin{figure}
\hspace{5 cm}\epsfxsize=230pt \epsfbox{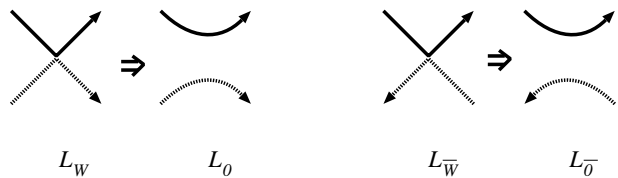}
\caption{Skein relations for intersections with kinks}
\label{figskein0}
\end{figure}

In all cases the operator $U(\gamma)$  deforms independently each line
at the intersection. For the cases with kinks that means that each
kink can be deformed into a smooth section. That means that in cases
$L_W$ and $L_{\bar{W}}$ the intersection can be removed by the 
deformation and we get the skein relations
\begin{eqnarray}
<W(L_W)> &= &<W(L_0)> \label{skeinkink1}\\
<W(L_{\bar{W})}> &=&<W(L_{\bar{0}})>  \label{skeinkink2}
\end{eqnarray}
that are shown in figure \ref{figskein0}

Notice that in all cases the removal of kinks at planar intersections 
can be accomplished given a straightforward  regularization of the
deformation operator. This provides a justification for the
identification of $L_W$ with $L_0$ that has been used in other 
works \cite{Br,MaSm}. We will return to discuss intersections with
kinks at the end of the next section, where we will analyze a
deformation that produces double lines in the loop that is of 
interest in the context of the exponentiation of the skein 
relation for straight through intersections, which we discuss now.

\section{Skein relations for intersections}

\subsection{Infinitesimal skein relations}
\begin{figure}
\hspace{6 cm}\epsfxsize=100pt \epsfbox{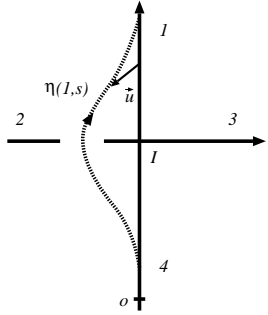}
\caption{Deformation of a ``straight through'' intersection. The
deformation vector $\vec{u}$ is perpendicular to the plane determined
by the intersection}
\label{intersecdeffig}
\end{figure}

In the case of a ``straight through'' intersection (no kinks),
the calculation is different than in the previously discussed cases.
In this case, the Wilson loop is the trace of the product of the
holonomies along the petals defined by the intersection. For instance
in the case of figure \ref{intersecdeffig} one could write it as
$W(\gamma) = {\rm Tr}(H_{12} H_{34})$.
The action of the loop derivative on the holonomy is independent of
the presence of the intersection, the expression is exactly the same
as (\ref{actionofloopder}). The difference in the construction arises
when one integrates by parts. Since the loop has a double point at the
intersection, the functional derivative with respect to $A_a$ has
two contributions at that point, corresponding to its  action on
the holonomy when it traverses that point the first and the second
time.  One of the contributions vanishes since it produces a term 
proportional to the tangent of the loop in the same plane as the
deformation and therefore spans no volume. The other contribution is
the one of interest. It can be written as, 
\begin{eqnarray}
\sigma^{ab} \Delta_{ab}(\eta(t)_o^x) <W(\gamma)> &=& 
-{4 \pi i\over k}  \int dA \sigma^{ab} 
e^{{i k\over 4 \pi} S_{CS}}
\epsilon_{abc} \int_{\gamma_{23}} dy^c \delta^3(x-y)\label{28}\\
&&\times {\rm Tr}\left(\tau^lH(\eta(t)_x^o)
H_{34}(\gamma_o^I)H_{12}(\gamma_I^y) \tau^l H_{12}(\gamma_y^I) 
H_{34}(\gamma_I^o)H(\eta(t)_o^x)\right).\nonumber
\end{eqnarray}
where we have assumed that the origin $o$ of the loop is in the petal
$34$. We now use the Fierz identity (\ref{Fierz}) and compute the
integral of the $\delta$ operator,
\begin{equation}
\int_0^1 dt \delta_{\vec{u}}(t) <{\rm Tr}(H_{12}(\gamma)
H_{34}(\gamma))> = 
-{i \pi \over k} {\rm Tr}(H_{12}(\gamma) H_{34}(\gamma))
+{2i \pi \over k} {\rm Tr}(H_{12}(\gamma))  {\rm Tr}(H_{34}(\gamma))
\label{29}
\end{equation}

Which can be rewritten as 
\begin{equation}
\int_0^1 dt \delta_{\vec{u}}(t) <W(L_I)>=
-{i \pi \over k} <W(L_I)>
+{2i \pi \over k} <W(L_W)>.
\label{30}
\end{equation}

A couple of remarks are in order. First, there is a difference between
equations (\ref{29}) and (\ref{30}), since equation (\ref{29}) refers
to the whole loop and equation (\ref{30}) only to its
intersection. The former has information about the connectivity of the
loop, the latter does not. Therefore, if one is to claim that equation
(\ref{30}) follows from equation (\ref{29}) one should prove that it
does so for any connectivity. For the case of double intersections,
there are only two different possible connectivities. It is
straightforward to check that equation (\ref{30}) holds independently
of the connectivity chosen. Second, in spite of similarities to 
equation (\ref{elemasmenos}), equation (\ref{30}) is only valid to 
first order in $1/k$ and needs to be exponentiated to obtain the 
skein relation. Since equation (\ref{30}) shows that the action of 
$\delta$ mixes $L_I$ and $L_W$, in order to exponentiate it we need
to compute the action of the deformation operator on $L_W$. Notice
that this is a different deformation than the one computed in the
last section, which was coplanar with the intersection. We 
discuss of the resulting calculation at the end of this 
section. 

Before doing that computation, it is worthwhile observing that one
could combine the Fierz identity (\ref{Fierz}) with the following
identity, 
\begin{equation}
\epsilon^{AC} \epsilon_{BD} = \delta^A_B \delta^C_D- \delta^A_D
\delta^C_B
\end{equation}
to get,
\begin{equation}
\tau^k{}^A_B \tau^k{}^C_D = {1\over 4} (\delta^A_D \delta^C_B- 
\epsilon^{AC} \epsilon_{BD}),
\end{equation}
and using this expression in (\ref{28}) we get
\begin{equation}
\int_0^1 dt \delta_{\vec{u}}(t) <{\rm Tr}(H_{12}(\gamma)
H_{34}(\gamma))> = 
{i \pi \over k} {\rm Tr}(H_{12}(\gamma))  {\rm Tr}(H_{34}(\gamma))
+{i \pi \over k} {\rm Tr}(H_{12}(\gamma) H_{34}^{-1}(\gamma)).
\end{equation}

In contrast to (\ref{29}) this expression leads to different
first-order skein relation depending on the connectivity of the
knot. If the the connectivity is such that the original loop has a
single component, then the skein relation is,
\begin{equation}
\int_0^1 dt \delta_{\vec{u}}(t) <W(L_I)>=
{i \pi \over k} <W(L_0)>
+{i \pi \over k} <W(L_{\bar{0}})>.
\label{31}
\end{equation}
where the element $L_{\bar{0}})$ is defined in figure
\ref{skeinrelfig}. The interest of this skein relation stems from the
fact that the original definition of the Kauffman bracket \cite{DuKa}
was given in terms of relations of this kind. Moreover, Major and
Smolin \cite{MaSm} used this identity to derive the binor identity
from the Kauffman bracket. The original definition of the bracket
differs from the one we use here in a factor $(-1)$ to elevated to the
number of connected components of the loop, which fixes the
connectivity difference we encountered above.

\begin{figure}
\hspace{6 cm}\epsfxsize=150pt \epsfbox{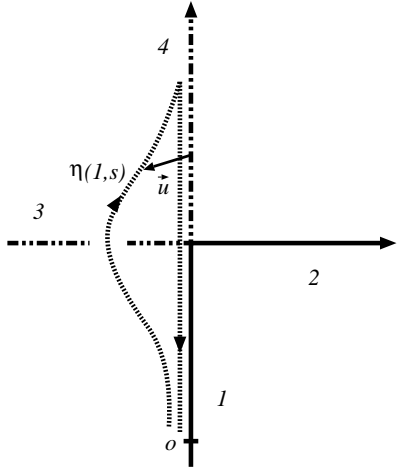}
\caption{The skein relation for intersections requires the analysis of
this type of deformation of an intersection with kinks.}
\label{deforrara}
\end{figure}
To conclude, we discuss the deformation of an $L_W$ as mentioned
before.
We need to compute the deformation of an intersection with a kink
that is shown in figure \ref{deforrara}. The loop starts at the
origin, goes through the deformation determined by the vector
$\vec{u}$, which we denote with a dotted line and then traverses the 
petal $12$ going through the kink and ends with the petal $34$. 
The resulting loop has double lines. In ordinary knot theory double
lines are not considered. They could be incorporated through
additional sets of skein relations, as we discuss in subsection F.
If one pursues a calculation similar to the ones we have been doing 
up to now for this case, one encounters two contributions, stemming
from the volumes spanned by the deformation and the tangents to the
loop at the intersection in lines $2$ or $3$. The integrals involve
terms of the form $\delta^3(y) \Theta(y)$. The Heaviside function
$\Theta(y)$
limits the integral in the contribution to the corresponding 
petal. Although these kinds of expressions can be regularized, there
does not appear any natural way of assigning relative weights to the
two contributions from the petals. The way we handle it here is 
to write the contribution in terms of two arbitrary factors $\alpha$
and $\beta$,
\begin{equation}
\int_0^1 dt <W(L_{W\pm})> =\pm {i \pi \over k} 
\left(\alpha <W(L_I)> + \beta <W(L_W)>\right).
\end{equation}
where we have denoted $W\pm$ the elements that result from doing
the deformation up or down with respect to the plane determined by
the intersection.

We will see in the next section that the arbitrary factors can be 
uniquely determined by the Mandelstam identities 
when we exponentiate the operator.

\subsection{Exponentiation and Mandelstam identities}

The skein relations involving $L_\pm$ and $L_I$ are obtained studying
the deformation of an intersection. In order to obtain these
deformations in a finite form we need to exponentiate the differential
expressions we obtained in the previous section. Specifically, we have
\begin{equation}
\left(\begin{array}{c}
      <W(L_\mp)>\\<W(L_{W\mp})>
      \end{array}\right) =
\exp\int_0^1 dt \delta_{\pm{\vec{u}}}(t) 
\left(\begin{array}{c}
      	<W(L_I)>\\
      	<W(L_W)>
      \end{array}\right) =
\exp\left(\pm {i \pi \over k}\right)
\left(\begin{array}{cc}
-1&2\\\alpha&\beta
\end{array}\right)
\left(\begin{array}{c}
      	<W(L_I)>\\
      	<W(L_W)>
      \end{array}\right). \label{skeinindet}
\end{equation}

We are interested only in computing $<W(L_\mp)>$. We are not
interested in the value of $<W(L_{W\mp})>$ since as we explained
before it involves double lines, which would require a separate
discussion involving extra skein relations (see subsection F). We do
not know the values of $\alpha$ and $\beta$. However, since we are
computing the invariant as an expectation value of a Wilson loop, we
expect it to satisfy the same relations that Wilson loops satisfy: the
Mandelstam identities.  We will see that imposing the Mandelstam
identities is enough to determine the coefficients $\alpha$ and
$\beta$ uniquely and therefore to find the skein relation.

\begin{figure}
\hspace{4 cm}\epsfxsize=300pt \epsfbox{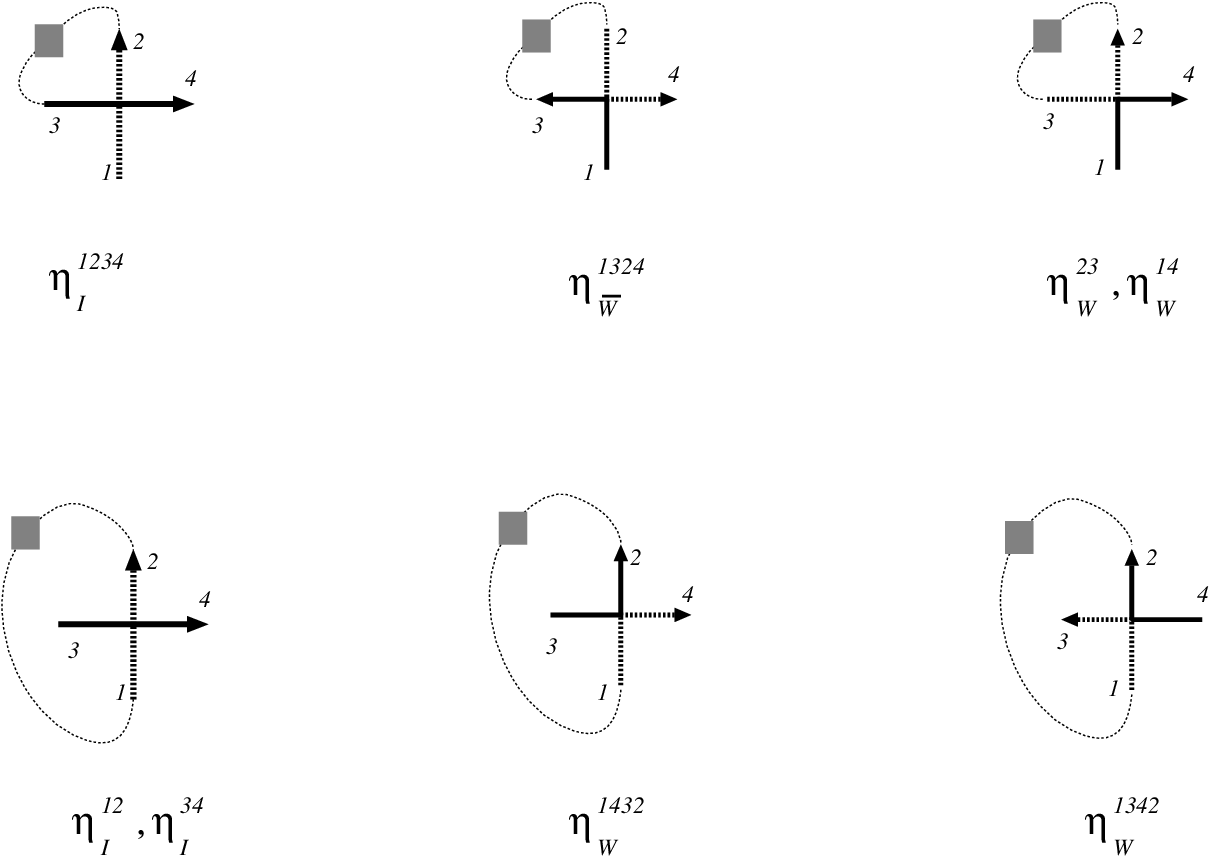}
\caption{The different loops that arise in the Mandelstam identities
involved in double intersections. The grey boxes denote possible
knottings and interlinkings with other portions of the loop (including
possible additional intersections and kinks). In order to unclutter
the figures, we only denoted partial connectivities, in all cases the
loops have to be closed by connecting the open strands, with no
particular restrictions, i.e. the resulting petal could have knottings
and interlinkings with the rest of the loop.}
\label{routingfig}
\end{figure}

\begin{figure}
\hspace{2 cm}\epsfxsize=250pt \epsfbox{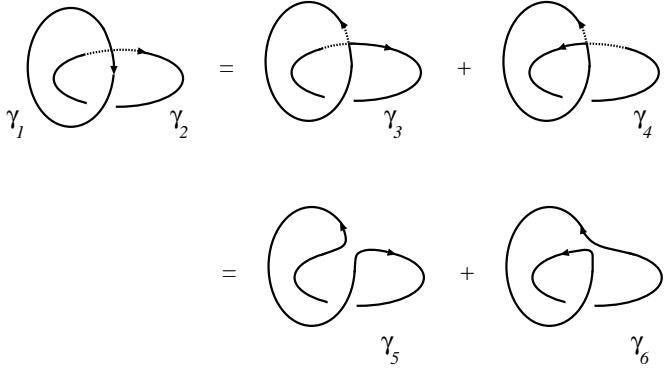}
\caption{Schematic depiction of the second identity 
used to determine $\alpha$ and $\beta$. }
\label{dosloopsfig}
\end{figure}

We therefore need to discuss the content of the Mandelstam
identities. As we mentioned before, these identities relate the
behavior of reroutings of the loops at intersections and involve
non-trivial information about the connectivity of the loop. We will
restrict the explicit discussion to double intersections, but it is
immediate to see that suitable generalizations can be found for more
complex intersections. For double intersections, one can always
consider a planar diagram, and two possible routings exist, as shown
in figure \ref{routingfig}. Each routing leads to an identity.
The identities,
\begin{eqnarray}
<W(\eta_I^{1234})> + <W(\eta_{\bar{W}}^{1324})> &=&
<W(\eta_W^{23}) W(\eta_W^{14})>\label{33}\label{34a}\\
<W(\eta_W^{1432})> + <W(\eta_{{W}}^{1342})> &=&
<W(\eta_I^{12}) W(\eta_I^{34})>,
\label{34}
\end{eqnarray}
involve different loops, as shown in the figure (\ref{routingfig}).
Moreover, they are valid for a completely arbitrary loop 
that has a double intersection. In order to use these
identities to allow us to fix the arbitrary parameters in the
exponentiation, we will consider ---for simplicity--- its expression
for a particular set of loops. We will then have to check that the
resulting invariant is consistent with the identities for all possible
loops.  The loops we wish to consider are the ones obtained by
reconnecting the loops of figure (\ref{routingfig}) with direct
strands, i.e. adding no knottings or interlinkings and are shown in
figures \ref{figureochofig} and \ref{dosloopsfig}. The Mandelstam
identities for these loops are,
\begin{figure}
\hspace{2 cm}\epsfxsize=300pt \epsfbox{fig11.eps}
\caption{Schematic depiction of the first identity used 
to determine $\alpha$ and $\beta$.}
\label{figureochofig}
\end{figure}
\begin{eqnarray}
<W(\eta_1)> +<W(\eta_2)> = <W(\eta_3) W(\eta_4)> \label{firstmand}\\
<W(\gamma_3)> +<W(\gamma_4)> = <W(\gamma_1) W(\gamma_2)>. \label{secondmand}
\end{eqnarray}

We now make use of the skein relations for intersections with kinks
(\ref{skeinkink1},\ref{skeinkink2}). In equation (\ref{firstmand})
this allows to replace $\eta_2$ by the unknot and $\eta_3$ and
$\eta_4$ by the two component unlinked link. Therefore $<W(\eta_2)>=1$.

In equation (\ref{secondmand}) the use of the skein relations is shown
in figure \ref{dosloopsfig} to transform $\gamma_3$ and $\gamma_4$ to
the unknot with a twist added. Using the skein relations
(\ref{skeintwist+},\ref{skeintwist-}) one gets, $<W(\gamma_5)> =q^{3
\over 4}$ and $<W(\gamma_6)> =q^{-{3\over 4}}$.

We now use (\ref{skeinindet}) in combination with these expressions to
determine the values of $\alpha$ and $\beta$. It turns out we only
need to use the expression of the exponential that appears in
(\ref{skeinindet}) expanded up to second order in $1/k$. Using this 
expression to determine $<W(\eta_1)>$, $<W(\gamma_1) W(\gamma_2)>$ and
$<W(\eta_3)W(\eta_4)>$ one gets that $\alpha=1$ and $\beta=0$. 
This completely characterizes the skein relation for the
intersections (\ref{skeinindet}). From there we conclude that,
\begin{equation}
\left(\begin{array}{c}
      <W(L_\mp)>\\<W(L_{W\mp})>
      \end{array}\right) =
\left(\begin{array}{cc}
q^{\pm{1\over 4}}&(q^{\mp{1\over 4}}-q^{\pm{1\over 4}})\\0&q^{\mp{1\over 4}}
\end{array}\right)
\left(\begin{array}{c}
      	<W(L_I)>\\
      	<W(L_W)>
      \end{array}\right). 
\end{equation}

From this expression, adding an subtracting with appropriate weights
the expressions with the different signs, we can work out the usual
skein relation for the Kauffman bracket without intersections,
\begin{equation}
q^{{1\over 4}} <W(L_+)> -q^{-{1\over 4}} <W(L_-)>
=(q^{{1\over 2}} -q^{-{1\over 2}}) <W(L_0)>,
\label{skein+-}
\end{equation}
and also the definition of the expectation value for a planar straight
through intersection,
\begin{equation}
<W(L_I)> = {1\over(q^{{1\over 4}}+q^{-{1\over 4}})} [<W(L_+)> +<W(L_-)>].
\label{skeinint}
\end{equation}

We therefore conclude that the expectation value of the Wilson loop in
an $SU(2)$  Chern--Simons theory is, up to a factor of 2 with the
conventions of this paper, identical to the Kauffman bracket knot
polynomial. 

\subsection{Consistency for all loops and to all orders in $1/k$}

We have just shown that considering the exponential of the
infinitesimal deformation to second order in $1/k$ and requiring
consistency with the Mandelstam identity for a particular set of loops
uniquely fixes the values of the indeterminate coefficients $\alpha$
and $\beta$ of the infinitesimal deformation. We need to check that
the construction is consistent to all orders in $1/k$ and for all
possible sets of loops with planar intersections. We will now show
that this is the case. In order to do this, the finite version of the
form of the skein relation (\ref{31}), which again can be derived in
the same form as the finite skein relation we derived above is useful,
\begin{equation}
<W(L_\pm)> = q^{\pm{1\over 4}} <W(L_0)> - q^{\mp{1\over 4}} <W(L_{\bar{0}})>. 
\label{kauorig}
\end{equation}
As we mentioned before, this relation depends on the connectivity of
the loop, we are assuming it is such that $L_I$ has one independent
component, as shown in loop $\eta_I^{1234}$ of figure
(\ref{routingfig}). Notice that by subtracting the $+$ and $-$ sign
versions of (\ref{kauorig}) we obtain (\ref{skein+-}). The sum contains
the information needed to derive the Mandelstam identity from the
skein relations. Combining (\ref{skeinint}) with (\ref{kauorig}) and
(\ref{skeinkink1},\ref{skeinkink2}) we get,
\begin{equation}
<W(L_I)> +<W(L_{\bar{W}})>= <W(L_W)>.
\label{I0W}
\end{equation}

Notice that we are stretching the notation here, since this identity
is not purely local, a given connectivity of the loop was assumed to
derive it, as mentioned above. Therefore one has to include the
connectivity we mentioned above in its interpretation. With that
connectivity, equation (\ref{I0W}) is exactly the Mandelstam
identity (\ref{34a}). This completes the proof. One could have chosen a
different connectivity when deriving (\ref{kauorig}) and then one
would have arrived at the Mandelstam identity (\ref{34}).

\subsection{Non-planar intersections}

Traditionally, in knot theory invariants have been formulated through
planar projections of knots. When generalizations to intersecting
knots are considered one may need to consider non-planar intersections
as inequivalent. This has already been noticed for triple
intersections \cite{ArUgGaMo}. In this section we will discuss some
skein relations satisfied by non-planar intersections. There are many
types of non-planar intersections, we will restrict ourselves to some
examples that have three of the four strands at the intersection in a
single plane. We will show that there exists a one-parameter family of
regularizations of the expectation value of the Wilson loop that is
compatible with the Mandelstam identities for the case of non-planar
double intersections. For one particular value of the parameter the
intersections behave as planar ones. For other values of the parameter
the intersections acquire common elements with under- and
over-crossings and therefore imply the existence of distinct
nontrivial generalizations of the usual invariants for the case on
non-planar intersections. 

\begin{figure}
\hspace{4 cm}\epsfxsize=200pt \epsfbox{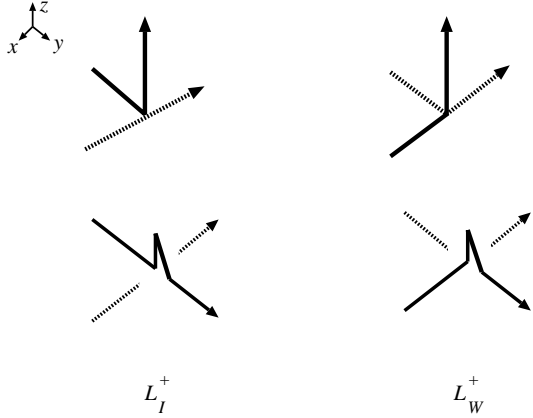}
\caption{The non-planar intersections we consider. The top row are
three dimensional views and the bottom row are the associated 
crossings that arise in the skein relations.}
\label{nonplanarfig}
\end{figure}

We consider a non-planar intersection as shown in figure
\ref{nonplanarfig}, which we represent with two new types of
crossings, labelled $L^{+}_I$ and $W^+_I$ with obvious counterparts
with a minus sign.

\begin{figure}
\hspace{6 cm}\epsfxsize=120pt \epsfbox{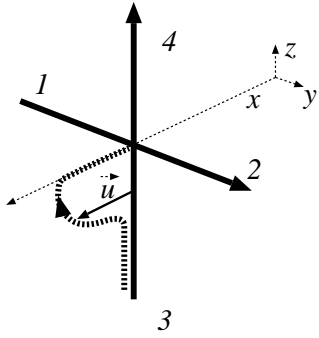}
\caption{The deformation of a straight through planar intersection
into a crossing  of the type $L^+_I$.}
\label{defornoplana}
\end{figure}

In order to compute the skein relations we consider a deformation of a
straight-through planar intersection as shown in figure
\ref{defornoplana}. The resulting integral is exactly the same as
equation (\ref{28}). The difference comes from the vector $\vec{u}$.
That vector vanishes in the $z>0$ part of the loop. This implies that 
when one wants to rederive equation (\ref{29}) one encounters 
ambiguities of the type $\delta(y) \Theta(y)$ exactly as when we 
deformed an $L_W$ intersection in a direction perpendicular to the
plane. The way to handle this is again to introduce an indeterminate
parameter $\alpha$. The resulting expression then reads,
intersections with 
\begin{equation}
\int_0^1 dt \delta_{\vec{u}}(t) <{\rm Tr}(H_{12}(\gamma)
H_{34}(\gamma))> = -\alpha [
{i \pi \over k} {\rm Tr}(H_{12}(\gamma) H_{34}(\gamma))
-{2i \pi \over k} {\rm Tr}(H_{12}(\gamma))  {\rm Tr}(H_{34}(\gamma))]
\end{equation}
this can again be reinterpreted as,
\begin{equation}
\int_0^1 dt \delta_{\vec{u}}(t) <W(L_I)>=\alpha [
{2i \pi \over k} <W(L_W)>
-{i \pi \over k} <W(L_I)>
].
\end{equation}

In order to exponentiate, since the action mixes  $L_I$ and $L_W$ we
need to compute,
\begin{equation}
\left(\begin{array}{c}
      <W(L^\mp_I)>\\<W(L^\mp_{W})>
      \end{array}\right) =
\exp\int_0^1 dt \delta_{\pm{\vec{u}}}(t) 
\left(\begin{array}{c}
      	<W(L_I)>\\
      	<W(L_W)>
      \end{array}\right) =
\exp\left(\pm {i \pi \over k}\right)
\left(\begin{array}{cc}
-\alpha &2\alpha\\\beta_1&\beta_2
\end{array}\right)
\left(\begin{array}{c}
      	<W(L_I)>\\
      	<W(L_W)>
      \end{array}\right)
\end{equation}
and one can show that in order to have consistency with the Mandelstam
identity, $\beta_1=0$ $\beta_2=\alpha$. Notice that $\alpha$ is
undetermined by the Mandelstam identities and we therefore have a
one-parameter family of definitions of the skein relation for the
crossing. Exponentiating explicitly and identifying the variable $k$
as before, we get,
\begin{eqnarray}
<W(L^\pm_I)> &=& q^{\mp{\alpha\over 4}} <W(L_I)>  +
(q^{\pm{\alpha\over 4}} -
q^{\mp{\alpha\over 4}})<W(L_0)>  \label{49}\\
<W(L^\pm_W)> &=& q^{\pm{\alpha\over 4}} <W(L_0)>.
\end{eqnarray}

If $\alpha=0$ we get that $L^\pm_I\equiv L_I$ and $L^\pm_W\equiv L_W$,
so the non-planar and planar intersections are treated in the same
way.

It is also immediate to prove from the skein relations
(\ref{49},\ref{skein+-},\ref{skeinint}) that,
\begin{eqnarray}
<W(L^\pm_I)>|_{\alpha=1} &=&<W(L_+)>\\
<W(L^\pm_I)>|_{\alpha=0} &=&<W(L_I)>\\
<W(L^\pm_I)>|_{\alpha=-1} &=&<W(L_-)>
\end{eqnarray}
so we see that for different values of the free parameter associated
with the non-planar intersections we can have them play the same role
as over-crossings, intersections and under-crossings. This highlights
the relation between the value of the $\alpha$ parameter and the
different types of regularizations it implies for the intersections.

As in the non-intersecting case, all the regular invariant information
of the $<W(\gamma)>$ is concentrated in a multiplicative ``phase
factor''. One can divide by it and construct an ambient isotopic
invariant, which for $SU(2)$ is the Jones polynomial. In the
non-intersecting case,  the writhe $w(\gamma)$ can be computed by
evaluating the expectation value of a Wilson loop in a $U(1)$
Chern--Simons theory, 
\begin{equation}
<W(\gamma)>|_{U(1)}= q^w(\gamma),
\end{equation}
with skein relations,
\begin{eqnarray}
w(\hat{L}_\pm) &=& \pm 1\\
w(L_\pm) &=& \pm 1\\
w(L_I) &=& w(L_0)= 0\\
w(L^\pm_I) &=& \pm \alpha
\end{eqnarray}

If one now defines a polynomial $J(\gamma)$ through
\begin{equation}
J(\gamma) = {1\over 2} q^{-{3\over 4}w(\gamma)} <W(\gamma)>,
\end{equation}
the result is ambient isotopic invariant. 

The above definition of the Jones polynomial is also valid for
multiloops, with a suitable generalization of the Abelian calculation
of the writhe for multiloops. From the expression of the resulting
Jones polynomial one can get an expression for the Gauss linking
number of two loops with intersections. The resulting expression is
consistent with lattice definitions of the linking number with
intersections \cite{Po,FoGaPu}.

\subsection{Triple and higher intersections}

We will not discuss in detail the generalization to triple
intersections in this paper, we sketch in this subsection how does one
perform the generalization of the construction to that case.
In the case of triple intersections, there are many types of possible
independent vertices. In the case of planar intersections it is easy
to see that one can deform them to double intersections, using the
same techniques as for the $L_W$'s. For the case of non-planar triple
intersections, there are 10 independent vertices. All of them can be
related to double intersections through deformations similar to the
one that connects $L_I$ with $L_\pm$. In order to exponentiate the
infinitesimal deformations, one again has to exponentiate a matrix 
that connects the different intersection types. It will be a sparse
$10\times 10$ matrix. Again regularization issues will leave many
coefficients undetermined and one would restrict them using the
Mandelstam identities. It is not clear if the resulting polynomial
will be completely determined by the Mandelstam identities or if new
free parameters will appear. As in the double case, the Mandelstam
identities are non-local and there are now three different possible
connectivities of intersections that are needed to implement the
identities. 

\subsection{Loops with multiple lines}

Throughout this paper we have assumed we were dealing with loops that
have each line traversed only one. If one would like the polynomial
that is being derived to take values on the complete set of loops that
is of interest in quantum gravity and gauge theories, one needs to
consider the case of loops with multiply traversed section. This is
also of importance if one is to view the integral of the exponential
of the Chern--Simons form as a rigorous measure on the space of
connections modulo gauge transformations. We present here a brief
discussion of how could one consider double lines, but a complete
description again requires further study. 

\begin{figure}
\hspace{5 cm}\epsfxsize=180pt \epsfbox{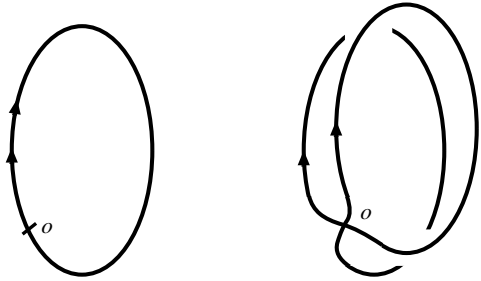}
\caption{The deformation of a loop traversed twice into two loops with
an intersection.}
\label{deforcuadro}
\end{figure}

Let us consider a loop $\gamma$ and study the Wilson loop along
$\gamma^2\equiv \gamma \circ \gamma$. Consider a generic direction in
space $\vec{w}$ such that the tangent vector to $\gamma$ is never
parallel to $\vec{w}$. Consider an continuous infinitesimal
displacement of all the points of $\gamma$ along $\vec{w}$ such that
an arbitrary point on $\gamma$ is kept fixed (we will call this point
the origin of the loop). This produces a second copy of $\gamma$.
If one deforms back and forth along $\vec{w}$ in this way one ends up
with two copies of $\gamma$ that are connected at the origin through
an intersection of the type $L_I$, as shown in figure
\ref{deforcuadro}. Because this is a planar deformation, the value of
the expectation value of the Wilson loop does not change. We have
therefore reduced the problem of computing the expectation value for a
loop traversed twice to the problem of computing the expectation value
along a simply traversed loop with an intersection of one of the types
studied before. It is worthwhile noticing that applying the Mandelstam
identity (\ref{34a}) at the resulting intersection one gets the
identity $<W(\gamma) W(\gamma)> = <W(\gamma\circ\gamma)> +2$.

If one wants to consider a more general situation, in which a loop
could have portions that are traversed twice, one can extend the above
result in a relatively straightforward manner. Even if there are
planar intersections of the multiply traversed segment with another
segment one can separate them without additional
contributions. However, if there are triple non-planar intersections
involving sections multiply traversed, the deformation will give
non-trivial contributions. These contributions can be evaluated
explicitly given a specific loop.

\section{Conclusions}

We have shown how to use variational techniques to obtain exact
expressions for the knot invariant associated with the expectation
value of the Wilson loop in Chern--Simons theory. The method is
completely general in the sense that it can be used in Chern--Simons
theory with any semisimple gauge group with small generalizations.  We
have worked out explicitly the value of the invariant not only for
smooth loops but also for loops with double planar and non-planar
intersections. The resulting invariant is compatible with the
Mandelstam identities of the gauge group and therefore is suitable for
providing invariants of interest as quantum states of topological
field theories and quantum gravity. Having a well defined 
linear function on the space of loops compatible with the Mandelstam
identities may allow also, using the techniques of Ashtekar and
collaborators \cite{AsLeMaMoTh} to define in a rigorous way a measure
in the space of connections modulo gauge transformations
$d\mu(A)$. Such a measure would allow to give rigorous meaning in a
mathematical sense to expression of the form $\int dA e^{kS_{CS}} f(A)$
for any gauge invariant function $f(A)$. 
The generalization of the work described in this paper to
triple intersections is straightforward and relevant for quantum
gravity applications. The possibility of computing explicitly the knot
polynomials associated with Chern--Simons theory for any group is
clearly of relevance in other physics applications, like the recent
discovery of Chern--Simons states in supergravity has shown
\cite{ArUgGaObPu}. It is expected that these results will be
extendible to $N>1$ supergravity where this method will provide with a
simple method of characterizing the potentially new invariants that
may arise.

\acknowledgements

We wish to thank Abhay Ashtekar, Leonardo Setaro and
Daniel Armand-Ug\'on for discussions.  This work was supported in part
by grants NSF-INT-9406269, NSF-PHY-9423950, NSF-PHY-9396246, research
funds of the Pennsylvania State University, the Eberly Family research
fund at PSU and PSU's Office for Minority Faculty development. JP
acknowledges support of the Alfred P. Sloan foundation through a
fellowship. We acknowledge support of Conicyt and PEDECIBA (Uruguay).

\appendix
\section{Commutativity of the $\delta$ operators}

We wish to evaluate the successive action of two $\delta$ operators on
$<W(\gamma)>$. We only discuss the case of a regular (non-intersecting)
point of the loop. The first $\delta$  acts as indicated in 
(\ref{actionofdeltat}), and using the Fierz identity we get,
\begin{equation}
\delta_{\vec{u}}(t) <W(\gamma)> = -{3 \pi i\over k} \int_0^1 ds
\dot{\eta}^a(s,t) {\eta'}^b(s,t) \epsilon_{abc}
\oint_\gamma dy^c \delta^3(\eta(s,t)- y) <W(\gamma)>.
\end{equation}

The second $\delta$ has two contributions, stemming from the action of
the loop derivative on the loop dependence of $<W(\gamma)>$ and
$\oint_\gamma$ respectively. The term that contributes to the commutator
is the latter, since the contribution on the Wilson loop is the same
in both orders. Let us therefore evaluate explicitly the non-trivial
contribution,
\begin{eqnarray}
[\delta(t_1),\delta(t_2)] &=& 
-{3 \pi i\over k} \int_0^1 ds_1 \int_0^1 ds_2 \dot{\eta}^a(s_1,t_1)
{\eta'}^b(s_1,t_1) \dot{\eta}^c(s_2,t_2) {\eta'}^d(s_2,t_2) 
\epsilon_{ab[c} \partial_{d]} \delta^3(x-\eta(s_1,t_1))|_{x=\eta(s_2,t_2)}
\nonumber\\
&&- (t_1\leftrightarrow t_2)\\
&=&-{3 \pi i\over k} \int_0^1 ds_1 \int_0^1 ds_2 \dot{\eta}^a(s_1,t_1)
{\eta'}^b(s_1,t_1) \dot{\eta}^c(s_2,t_2)
\epsilon_{abc}\partial_{s_2} \delta^3(\eta(s_2,t_2)-\eta(s_1,t_1))\nonumber\\
&&+{3 \pi i\over k} \int_0^1 ds_1 \int_0^1 ds_2 \dot{\eta}^a(s_1,t_1)
{\eta'}^b(s_1,t_1)  {\eta'}^d(s_2,t_2) 
\epsilon_{abd} \partial_{t_2} \delta^3(\eta(s_2,t_2)-\eta(s_1,t_1))
\nonumber\\
&&- (t_1\leftrightarrow t_2).
\end{eqnarray}

One can now pick a coordinate chart in which 
$\delta^3(\eta(s_2,t_2)-\eta(s_1,t_1))=
Z \delta(s_2-s_1) \delta(t_2-t_1)$, where
$Z$ is a regularization dependent factor that does not depend on $s_2$ or
$t_2$. Using the distributional identity,
\begin{equation}
f(y) \partial_y \delta(y-x) =f(x) \partial_y 
\delta(y-x)- \delta(y-x) \partial_x f(x)
\end{equation}
it is immediate to check that the contributions vanish. The result is
obviously dependent on a regularization choice.

\end{document}